\documentclass[pra,aps,amsmath,amssymb,twocolumn,longbibliography]{revtex4-1}
\usepackage{graphicx,multirow}
\usepackage{color,outlines}
\usepackage[sc,osf]{mathpazo}
\usepackage{amsmath}

\usepackage{physics}
\usepackage{outlines}

\usepackage{hyperref}
\usepackage{dcolumn}
\usepackage{bm}
\usepackage{color}
\usepackage{xcolor}
\usepackage{bbold}
\usepackage{soul}
\usepackage{epsfig,epic}
\usepackage{epstopdf}
\usepackage{amsthm}

\usepackage{cleveref}
\usepackage{algorithm2e}
\newcommand{\blue}{\color{blue}}

\newcommand{\newc}{\newcommand}
\newc{\beq}{\begin{equation}}
\newc{\eeq}{\end{equation}}
\newc{\kt}{\rangle}
\newc{\br}{\langle}
\newc{\beqa}{\begin{eqnarray}}
\newc{\eeqa}{\end{eqnarray}}
\newc{\ovl}{\overline}
\usepackage[flushleft]{threeparttable} 
\usepackage{tabularx}
\usepackage{array}
\usepackage{soul}
\usepackage{nicefrac}

\usepackage[T1]{fontenc}
\usepackage{amsmath,amsfonts,amssymb,amsthm,bbm,bm, braket}
\usepackage{wasysym}
\usepackage{comment}
\usepackage{scalerel}
\usepackage{enumerate}
\usepackage{graphicx}
\usepackage{mathtools}
\usepackage{ifthen}
\usepackage{tensor}
\usepackage{tikz}
\usepackage{tikz-network}
\usetikzlibrary{patterns,decorations.pathreplacing}

\theoremstyle{break}

\usepackage{color}
\definecolor{myred}{RGB}{232,102,102}
\definecolor{myblue}{RGB}{150,150,255}
\definecolor{myorange}{RGB}{255,165,0}
\definecolor{mygrey}{RGB}{105,105,105}
\definecolor{OliveGreen}{RGB}{85,107,47}
\definecolor{NavyBlue}{RGB}{0,0,128}
\definecolor{mygreen}{RGB}{34,139,34}
\definecolor{myY}{RGB}{220,255,203}
\definecolor{myYO}{RGB}{255, 220, 151}
\usepackage{tensor}
\newcommand{\be}{\begin{equation}}
\newcommand{\ee}{\end{equation}}
\newcommand{\ba}{\begin{aligned}}
\newcommand{\ea}{\end{aligned}}
\newcommand{\bw}{\begin{widetext}}
\newcommand{\ew}{\end{widetext}}

\theoremstyle{plain}

\theoremstyle{plain}

\begin{document}

\title{ Solvable models of many-body chaos from dual-Koopman circuits} 
\author{Arul
Lakshminarayan}
\email{arul@physics.iitm.ac.in} 
\affiliation{Department of Physics, Indian Institute of Technology
Madras, Chennai, India~600036}
   
\begin{abstract}
Dual-unitary circuits are being vigorously studied as models of many-body quantum chaos that can be solved exactly for correlation functions and time evolution of states. Here we study their classical counterparts defining dual-canonical transformations and associated dual-Koopman operators. Classical many-body systems constructed these have the property, like their quantum counterparts, that the correlations vanish everywhere except on the light-cone, on which they decay with rates governed by a simple contractive map. Providing a large class of such dual-canonical transformations, we study in detail the example of a coupled standard map and show analytically that arbitrarily away from the integrable case, in the thermodynamic limit the system is mixing. We also define ``perfect" Koopman operators that lead to the correlation vanishing everywhere including on the lightcone and provide an example of a cat-map lattice which 
would qualify to be a Bernoulli system at the apex of the ergodic hierarchy.

\end{abstract}

\maketitle

Research on quantum many-body systems has recently seen an explosion, both in theoretical formalism and in controlled experiments, for example see \cite{Browaeys_2020,Monroe_2021} and references therein. Riding on the fast developments in quantum information and computing technologies, they additionally present interesting test beds for the study of quantum chaos \cite{Haake_2013,Alessio_2016}. A class of one-dimensional systems, dubbed dual-unitary circuits \cite{Bertini_2019} has been particularly attractive as they could be strongly nonintegrable, in the sense of having random matrix fluctuations, and yet solvable in some important ways. This exploits a duality between space and time directions in the elementary building blocks of the nearest neighbor particles. A rapidly growing body of work explore dynamics of these quantum dual unitary systems  \cite{Akila_2016,Sarang_2019,Bertini_2018,Bertini_2019,Gutkin_2020,Claeys_2020,Lorenzo_2020,
Claeys_2021,ASA_2021,Kos_2021,Bertini_2021CMP,Jonay_2021,Alessio_2021,Tianci_2022,Gombor_2022,Suhail_2022,Claeys_2022,Rampp_2023,Foligno_2023,Kasim_2023}. 
The most important advantage of duality is that it enables the exact evaluation of otherwise inaccessible correlation functions, especially in the thermodynamic limit.
Apart from some Floquet spin chains though, these dual-unitary circuits are arguably 
abstract models with no apparent relation, if a  classical limit exists, to Hamiltonian chaos. For example, are there dual-unitary circuits whose classical limit  can apply the Kolmogorov-Arnold-Moser (KAM) theorem and the associated homoclinic tangle route to chaos \cite{Ott_book,KAM_book}? 

In this work, we start from purely classical models and ask for properties that lead to space-time duality. Quantization of these models are a subset of dual-unitary circuits with semiclassical limits. Apart from that, spatiotemporal chaos in classical extended systems and fields is in itself of great interest, for example see \cite{Kaneko_1989,Liang_2022,Gutkin_2021, Lando_2023}. A pioneering work has studied ``particle-time" duality
in classical settings \cite{Gutkin_2016}, however, the general characterization of the classical equivalence of dual unitarity seems to be absent. For example, in the thermodynamic limit, dual-unitary circuits are known to have vanishing single-particle correlations everywhere except on the light-cone $x=\pm t$. Classical correlations can also behave similarly under conditions of duality. Notably, we can study mixing in the 
thermodynamic limit of coupling near-integrable systems. In an explicit example it is shown below that in such a limit, even infinitesimal breaking of integrability can lead to mixing characterized by exponential decay of single particle correlations.

It turns out to be natural to start with the Hilbert space approach to classical mechanics due to Koopman \cite{Koopman_1931} who was inspired by its fundamental appearance in quantum mechanics. The Koopman operator $\hat K$ evolves functions in phase space rather than individual points. Thus if $\Phi$ represents a Hamiltonian flow over some time $t$, and if $x$ is a phase space point, then for a function $f(x)$, $\hat{K}f(x)=f(\Phi(x))$. Much of dynamical systems theory and the developments in chaos though happened following phase space points of the orbit $x \mapsto \Phi(x)$. However, such orbits are not very useful in higher dimensional, or many-body, systems.
 
The Koopman operator is infinite dimensional for finite degrees of freedom, but redeems itself by being a linear unitary operator, much like the quantum propagator. In the past two decades, the Koopman formalism has seen a remarkable resurgence with applications in nonlinear control systems and machine learning of dynamical processes \cite{Koopman_book}. The present work aspires to add to the prevailing ``Koopmania" \cite{ChaosBook} by pointing out that there are dual-unitary Koopman (or simply dual-Koopman) operators which can be used to construct, in some aspects, solvable many-body chaotic systems.

 Apart from dual-Koopman operators we also point to the existence of the classical equivalents of multiunitary \cite{Goyeneche_2015} or perfect-tensors \cite{Pastawski_2015,Hosur_2016} which have additional duality properties. In the quantum case, this enables the construction of absolutely maximally entangled states \cite{Goyeneche_2015,Rather_2022} and renders the many-body systems constructed from them Bernoulli \cite{Ergodic_Hierarchy}. Bernoulli systems occupy the apex of the ergodic hierarchy, and have everywhere vanishing correlation functions, including on the light-cone \cite{ASA_2021}. Classical ``perfect Koopman" operators  are similarly defined below and explicit examples are displayed.

 Tensor network methods that vastly aid computations in quantum many-body systems have a very natural, and simpler translation to the Koopman operators constructing these classical systems. In the quantum circuits, the unitary gates $U$ are usually two-particle propagators and the evolution of operators is the Heisenberg adjoint action $\hat{a}\rightarrow U^{\dagger} \hat{a} U$, whose direct action is via vectorization of the operators: $\ket{\hat{a}} \rightarrow (U^{\dagger}  \otimes U^T) \ket{\hat{a}}$.
Thus in the corresponding classical circuits the Koopman operator $\hat{K}$
is the equivalent of  $U^{\dagger}  \otimes U^T$ rather than $U$ itself. In terms of circuits the natural classical ones then correspond to the so-called ``folded" quantum circuits rather than the ones where the propagators themselves figure.

Consider a canonical transformation $(q_1,p_1,q_2,p_2) \mapsto (q'_1,p'_1,q'_2,p'_2)$ in two-degrees of freedom. Let ``new" variables  $\xi_1'=(q_1',p_1')=\phi_1(\xi_1,\xi_2)$ and $\xi_2'=(q_2',p_2')=\phi_2(\xi_1,\xi_2)$ be obtained from the ``old" ones $\xi_1=(q_1,p_1)$ and $\xi_2=(q_2,p_2)$. This defines a
four dimensional symplectic map on phase space $\Omega$ whose iterations form a dynamical system. For example the Poincar\'e surface of section of a three-degree of freedom Hamiltonian flow, or the stroboscopic map of a two-degree of freedom periodically forced system is such a map. 
To concentrate on the essentials, we will assume the phase-spaces to be compact, and later give concrete examples where they are tori. In addition we assume that the units are such that the phase-space area of the individual spaces and the combined ones are unity.

The Koopman operator's matrix elements or kernel is
\beq
\begin{split}
&\bra{\xi_1',\xi_2'}\hat{K}\ket{\xi_1,\xi_2} \equiv K(\xi_1',\xi_2';\xi_1,\xi_2)=\\
&\delta\left[\xi_1-\phi_1(\xi_1',\xi_2'\right)]
\delta\left[ \xi_2-\phi_2(\xi_1',\xi_2')\right].
\end{split}
\label{eq:Koopman}
\eeq

This can be represented diagrammatically as  
\be
\begin{tikzpicture}[baseline=(current  bounding  box.center), scale=1]
\def\eps{0.5};
\draw[thick] (-4.25,0.5) -- (-3.25,-0.5);
\draw[thick] (-4.25,-0.5) -- (-3.25,0.5);
\draw[ thick, fill=myblue, rounded corners=2pt] (-4,0.25) rectangle (-3.5,-0.25);
\draw[thick] (-3.75,0.15) -- (-3.6,0.15) -- (-3.6,0);
\Text[x=-4.35,y=-0.65, anchor = center]{$\xi_1$}
\Text[x= -3.10,y=-0.65, anchor = center]{$\xi_2$}
\Text[x=-4.35,y=0.7, anchor = center]{$\xi'_1$}
\Text[x=-3.10,y=0.7, anchor = center]{$\xi'_2$}
\end{tikzpicture}
\label{eq:Koopman_diag}
\ee
where the particles are in the horizontal direction and time vertical.  The legs are labeled by the phase-space coordinates of a given degree of freedom and a line 
$ \xi\, \begin{tikzpicture} \draw[thick] (0,0) -- (.75,0); \end{tikzpicture} \xi'$ 
is the Dirac delta function $\delta(\xi-\xi')$, and the square represents the Koopman operator. An arbitrary function on a single particle phase-space, say $F(\xi)$, is represented by a small black filled circle $\xi \, \begin{tikzpicture} \draw[thick] (0,0) -- (.75,0); \draw[thick, fill=black] (.75,0) circle (0.075cm); \end{tikzpicture}$.

The unitarity of $\hat{K}$ follows from the transformations being canonical.
However, the more critical property is ``unitality" \cite{Kos_2021}.  If $F_u(\xi)=1$ be the constant function, we have
\beq
\begin{split}
\int_{\Omega} K(\xi_1',\xi_2';\xi_1,\xi_2) d\xi_1 d\xi_2 &=F_u(\xi_1') F_u(\xi_2'), \\
\int_{\Omega} K(\xi_1',\xi_2';\xi_1,\xi_2) d\xi_1' d\xi_2' &=F_u(\xi_1) F_u(\xi_2),
\end{split}
\label{eq:unitality_eqns}
\eeq
the first being a trivial consequence of Eq.~(\ref{eq:Koopman}), and the second follows from the uniform Liouville measure being invariant.
If the constant function $F_u(\xi)$ is represented by an unfilled circle, the above 
are diagrammatically
\be
\begin{tikzpicture}[baseline=(current  bounding  box.center), scale=1]
\def\eps{0.5};
\draw[thick] (-4.25,0.5) -- (-3.25,-0.5);
\draw[thick] (-4.25,-0.5) -- (-3.25,0.5);
\draw[ thick, fill=myblue, rounded corners=2pt] (-4,0.25) rectangle (-3.5,-0.25);
\draw[thick] (-3.75,0.15) -- (-3.6,0.15) -- (-3.6,0);
\Text[x=-2.75,y=0.0, anchor = center]{$=$}
\draw[thick, fill=white] (-4.25,-0.5) circle (0.075cm); 
\draw[thick, fill=white] (-3.25,-0.5) circle (0.075cm); 
\draw[thick] (-2.25,0.5) -- (-2.25,-0.5);
\draw[thick] (-1.25,-0.5) -- (-1.25,0.5);
\draw[thick, fill=white] (-2.25,-0.5) circle (0.075cm); 
\draw[thick, fill=white] (-1.25,-0.5) circle (0.075cm); 
\end{tikzpicture},
\begin{tikzpicture}[baseline=(current  bounding  box.center), scale=1]
\def\eps{0.5};
\draw[thick] (-4.25,0.5) -- (-3.25,-0.5);
\draw[thick] (-4.25,-0.5) -- (-3.25,0.5);
\draw[ thick, fill=myblue, rounded corners=2pt] (-4,0.25) rectangle (-3.5,-0.25);
\draw[thick] (-3.75,0.15) -- (-3.6,0.15) -- (-3.6,0);
\Text[x=-2.75,y=0.0, anchor = center]{$=$}
\draw[thick, fill=white] (-4.25,0.5) circle (0.075cm); 
\draw[thick, fill=white] (-3.25,0.5) circle (0.075cm); 
\draw[thick] (-2.25,0.5) -- (-2.25,-0.5);
\draw[thick] (-1.25,-0.5) -- (-1.25,0.5);
\draw[thick, fill=white] (-2.25,0.5) circle (0.075cm); 
\draw[thick, fill=white] (-1.25,0.5) circle (0.075cm); 
\end{tikzpicture},
\label{eq:unitality_diag}
\ee
where the phase-space variables are suppressed as they will be from now.

A dual is obtained when the roles of time and space are interchanged. The  
``old" variables are now $(q_1,p_1,q_1',p_1')$ and the ``new"  $(q_2,p_2,q_2',p_2')$. The original canonical transform will be ``dual-canonical" if this is also a canonical transformation. While we display explicit examples of such transformations below, the resulting conditions on the Koopman operator is of immediate interest and we represent them in terms of diagrams as
\be
\begin{tikzpicture}[baseline=(current  bounding  box.center), scale=1]
\def\eps{0.5};
\draw[thick] (-4.25,0.5) -- (-3.25,-0.5);
\draw[thick] (-4.25,-0.5) -- (-3.25,0.5);
\draw[ thick, fill=myblue, rounded corners=2pt] (-4,0.25) rectangle (-3.5,-0.25);
\draw[thick] (-3.75,0.15) -- (-3.6,0.15) -- (-3.6,0);
\Text[x=-2.75,y=0.0, anchor = center]{$=$}
\draw[thick, fill=white] (-3.25,-0.5) circle (0.075cm); 
\draw[thick, fill=white] (-3.25, 0.5) circle (0.075cm); 
\draw[thick] (-1.25,0.5) -- (-2.25, 0.5);
\draw[thick] (-1.25,-0.5) -- (-2.25,-0.5);
\draw[thick, fill=white] (-1.25, 0.5) circle (0.075cm); 
\draw[thick, fill=white] (-1.25,-0.5) circle (0.075cm); 
\Text[x=-1,y=0.0, anchor = center]{,}
\end{tikzpicture}
\qquad\,\, 
\begin{tikzpicture}[baseline=(current  bounding  box.center), scale=1]
\def\eps{0.5};
\draw[thick] (-4.25,0.5) -- (-3.25,-0.5);
\draw[thick] (-4.25,-0.5) -- (-3.25,0.5);
\draw[ thick, fill=myblue, rounded corners=2pt] (-4,0.25) rectangle (-3.5,-0.25);
\draw[thick] (-3.75,0.15) -- (-3.6,0.15) -- (-3.6,0);
\Text[x=-2.75,y=0.0, anchor = center]{$=$}
\draw[thick, fill=white] (-4.25,0.5) circle (0.075cm); 
\draw[thick, fill=white] (-4.25,-0.5) circle (0.075cm); 
\draw[thick] (-2.25,0.5) -- (-1.25,0.5);
\draw[thick] (-2.25,-0.5) -- (-1.25,-0.5);
\draw[thick, fill=white] (-2.25,-0.5) circle (0.075cm); 
\draw[thick, fill=white] (-2.25,0.5) circle (0.075cm);
\end{tikzpicture}.
\label{eq:dualunitality_diag}
\ee
The integration variables in the equivalent to Eq.~(\ref{eq:unitality_eqns}) will be the dual old and new coordinates.
A Koopman operator that satisfies Eqs.~(\ref{eq:unitality_diag},\ref{eq:dualunitality_diag}) is ``dual-Koopman", the classical equivalent of the dual-unitary operators of quantum circuits.

One can also define ``$\Gamma$-duality" (also referred to a T-duality in \cite{ASA_2021} for quantum gates) as the conditions that 
\be
\begin{tikzpicture}[baseline=(current  bounding  box.center), scale=1]
\def\eps{0.5};
\draw[thick] (-4.25,0.5) -- (-3.25,-0.5);
\draw[thick] (-4.25,-0.5) -- (-3.25,0.5);
\draw[ thick, fill=myblue, rounded corners=2pt] (-4,0.25) rectangle (-3.5,-0.25);
\draw[thick] (-3.75,0.15) -- (-3.6,0.15) -- (-3.6,0);
\Text[x=-2.75,y=0.0, anchor = center]{$=$}
\draw[thick, fill=white] (-3.25,-0.5) circle (0.075cm); 
\draw[thick, fill=white] (-4.25, 0.5) circle (0.075cm); 
\draw[thick] (-1.25,0.5) -- (-2.25, 0.5);
\draw[thick] (-1.25,-0.5) -- (-2.25,-0.5);
\draw[thick, fill=white] (-2.25, 0.5) circle (0.075cm); 
\draw[thick, fill=white] (-1.25,-0.5) circle (0.075cm); 
\Text[x=-1,y=0.0, anchor = center]{,}
\end{tikzpicture}
\qquad\,\, 
\begin{tikzpicture}[baseline=(current  bounding  box.center), scale=1]
\def\eps{0.5};
\draw[thick] (-4.25,0.5) -- (-3.25,-0.5);
\draw[thick] (-4.25,-0.5) -- (-3.25,0.5);
\draw[ thick, fill=myblue, rounded corners=2pt] (-4,0.25) rectangle (-3.5,-0.25);
\draw[thick] (-3.75,0.15) -- (-3.6,0.15) -- (-3.6,0);
\Text[x=-2.75,y=0.0, anchor = center]{$=$}
\draw[thick, fill=white] (-3.25,0.5) circle (0.075cm); 
\draw[thick, fill=white] (-4.25,-0.5) circle (0.075cm); 
\draw[thick] (-2.25,0.5) -- (-1.25,0.5);
\draw[thick] (-2.25,-0.5) -- (-1.25,-0.5);
\draw[thick, fill=white] (-1.25,0.5) circle (0.075cm); 
\draw[thick, fill=white] (-2.25,-0.5) circle (0.075cm);
\end{tikzpicture}.
\label{eq:Gammadualunitality_diag}
\ee
Note that in the above diagrams the horizontal lines on the right sides of the equations could well also be vertical ones.
If a dual-Koopman operator is also $\Gamma$-dual, it is the classical equivalent of a perfect tensor or 2-unitary gate \cite{Goyeneche_2015}, hence it can be justifiably referred to as  a ``perfect-Koopman operator". If this is the case, it has strong implications for many-body systems built from them as they will be be Bernoulli systems wherein correlations instantly vanish.

In general Eq.~(\ref{eq:dualunitality_diag}) or Eq.~(\ref{eq:Gammadualunitality_diag}) are not satisfied, and we can define four ``partially traced" Koopman maps:
\beq
\begin{split}
\bra{\xi_1'}\hat{K}_{R_1}\ket{\xi_1}=&
\int K\, d\xi_2 d\xi'_2, \; \bra{\xi_2'}\hat{K}_{R_2}\ket{\xi_2}=
\int K\, d\xi_1 d\xi'_1 \\
\bra{\xi_2'}\hat{K}_{\Gamma}^{+}\ket{\xi_1}=&
\int K\,d\xi_1' d\xi_2,\, \bra{\xi_1'}\hat{K}_{\Gamma}^{-}\ket{\xi_2}=
\int K\,d\xi_1 d\xi_2'
\end{split}
\label{eq:Maps}
\eeq 
where the integration range is over all the relevant spaces. If $\hat{K}$ is dual, then $\hat{K}_{R_i}=\ket{F_u}\bra{F_u}$  and if it is $\Gamma$-dual, then $\hat{K}_{\Gamma}^{\pm}=\ket{F_u}\bra{F_u}$, that is they are rank-1 projectors onto the uniform function. In general they are sub-unitary operators, but in any case they have $F_u(\xi)$ as a ``trivial" eigenfunction with eigenvalue $1$. In the corresponding quantum case they define completely positive trace-preserving maps, channels with dynamical significance for circuits constructed of them. In classical circuits they play exactly the same roles as we see below.

Let $\hat{K}_i$ and $\hat{K'}_i$  ($i=1,2$) be Koopman operators corresponding to two canonical transformations on the individual particle phase-spaces; {\it i.e.}, they are {\it local} Koopman operators. Let $\hat{K}_{12}$ be a Koopman operator corresponding to a general joint canonical transform
in 4-dim. phase space. Then $\hat{K}=(\hat{K'}_1 \otimes \hat{K'}_2) \hat{K}_{12} (\hat{K}_1 \otimes \hat{K}_2)$ is also a Koopman operator respecting unitality condition as in Eq.~(\ref{eq:unitality_diag}). This is proved in Eq.~(\ref{eq:localunitality_diag}), where the local Koopman operators are represented by blue circles, while $\hat{K}_{12}$ is still the square, and the unitality of the local Koopman operators is used.
\be
\begin{tikzpicture}[baseline=(current  bounding  box.center), scale=1]
\def\eps{0.5};
\draw[thick] (-4.75,.75) -- (-3.25,-.75);
\draw[thick] (-4.75,-.75) -- (-3.25,.75);
\draw[ thick, fill=myblue, rounded corners=2pt] (-4.25,-0.25) rectangle (-3.75,0.25);
\draw[thick] (-4,0.15) -- (-3.85,0.15) -- (-3.85,0);
\draw[thick, fill=myblue] (-4.5,-0.5) circle (0.125cm); 
\draw[thick, fill=myblue] (-3.5,-0.5) circle (0.125cm);
\draw[thick, fill=myblue] (-3.5, 0.5) circle (0.125cm); 
\draw[thick, fill=myblue] (-4.5, 0.5) circle (0.125cm);
\draw[thick, fill=white] (-4.75,-.75) circle (0.075cm); 
\draw[thick, fill=white] (-3.25,-.75) circle (0.075cm); 
\Text[x=-2.75,y=0.0, anchor = center]{$=$}
\end{tikzpicture}
\begin{tikzpicture}[baseline=(current  bounding  box.center), scale=1]
\def\eps{0.5};
\draw[thick] (-4.75,.75) -- (-3.25,-.75);
\draw[thick] (-4.75,-.75) -- (-3.25,.75);
\draw[ thick, fill=myblue, rounded corners=2pt] (-4.25,-0.25) rectangle (-3.75,0.25);
\draw[thick] (-4,0.15) -- (-3.85,0.15) -- (-3.85,0);
\draw[thick, fill=myblue] (-3.5, 0.5) circle (0.125cm); 
\draw[thick, fill=myblue] (-4.5, 0.5) circle (0.125cm);
\draw[thick, fill=white] (-4.75,-.75) circle (0.075cm); 
\draw[thick, fill=white] (-3.25,-.75) circle (0.075cm); 
\Text[x=-2.75,y=0.0, anchor = center]{$=$}
\draw[thick] (-2.25,0.75) -- (-2.25,-0.75);
\draw[thick] (-1.25,-0.75) -- (-1.25,0.75);
\draw[thick, fill=myblue] (-2.25, 0.5) circle (0.125cm); 
\draw[thick, fill=myblue] (-1.25, 0.5) circle (0.125cm);
\draw[thick, fill=white] (-2.25,-.75) circle (0.075cm); 
\draw[thick, fill=white] (-1.25,-.75) circle (0.075cm); 
\Text[x=-1.,y=0.0, anchor = center]{$=$}
\draw[thick] (-.5,0.75) -- (-.5,-0.75);
\draw[thick] (0.5,-0.75) -- (0.5,0.75);
\draw[thick, fill=white] (-.5,-.75) circle (0.075cm); 
\draw[thick, fill=white] (.5,-.75) circle (0.075cm); 
\end{tikzpicture}
\label{eq:localunitality_diag}
\ee
While this is just a consequence of a composition of canonical transforms being canonical, more importantly, if $\hat{K}_{12}$ is dual or $\Gamma$-dual or perfect,
then so is $\hat{K}$, these properties are preserved under local transformations. The proofs are very similar to that presented in Eq.~(\ref{eq:localunitality_diag}), for example duality relation is proved by just viewing this equation flow sideways. Note that the composition of dual-Koopman operators is in general not dual, and the nature of the composed canonical transforms being local is important.

A simple, but important dual-canonical transform is the ``swap", which interchanges the two particles: $(q_1',p_1',q_2',p_2')=(q_2,p_2,q_1,p_1)$. Denoting the corresponding Koopman operator $\hat{K}_s$ we have 
\be
\begin{split}
&\begin{tikzpicture}[baseline=(current  bounding  box.center), scale=1]
\def\eps{0.5};
\draw[thick] (-4.25,0.5) -- (-3.25,-0.5);
\draw[thick] (-4.25,-0.5) -- (-3.25,0.5);
\draw[ thick, fill=myred, rounded corners=2pt] (-4,0.25) rectangle (-3.5,-0.25);
\draw[thick] (-3.75,0.15) -- (-3.6,0.15) -- (-3.6,0);
\Text[x=-2.75,y=0.0, anchor = center]{$=$}
\draw[thick, fill=white] (-4.25,-0.5) circle (0.075cm); 
\draw[thick, fill=white] (-3.25,-0.5) circle (0.075cm); 
\draw[thick] (-2.25,0.5) -- (-1.8,.05);
\draw[thick] (-1.7,-0.05)-- (-1.25,-0.5);
\draw[thick] (-2.25,-0.5) -- (-1.25,0.5);
\draw[thick, fill=white] (-2.25,-0.5) circle (0.075cm); 
\draw[thick, fill=white] (-1.25,-0.5) circle (0.075cm); 
\Text[x=-.75,y=0.0, anchor = center]{$=$}
\draw[thick] (-.25,0.5) -- (-.25,-.5);
\draw[thick] (.75,0.5)-- (.75,-0.5);
\draw[thick, fill=white] (-.25,-0.5) circle (0.075cm); 
\draw[thick, fill=white] ( .75,-0.5) circle (0.075cm); 
\end{tikzpicture},\\
& \begin{tikzpicture}[baseline=(current  bounding  box.center), scale=1]
\def\eps{0.5};
\draw[thick] (-4.25,0.5) -- (-3.25,-0.5);
\draw[thick] (-4.25,-0.5) -- (-3.25,0.5);
\draw[ thick, fill=myred, rounded corners=2pt] (-4,0.25) rectangle (-3.5,-0.25);
\draw[thick] (-3.75,0.15) -- (-3.6,0.15) -- (-3.6,0);
\Text[x=-2.75,y=0.0, anchor = center]{$=$}
\draw[thick, fill=white] (-3.25,0.5) circle (0.075cm); 
\draw[thick, fill=white] (-3.25,-0.5) circle (0.075cm); 
\draw[thick] (-2.25,0.5) -- (-1.8,.05);
\draw[thick] (-1.7,-0.05)-- (-1.25,-0.5);
\draw[thick] (-2.25,-0.5) -- (-1.25,0.5);
\draw[thick, fill=white] (-1.25, 0.5) circle (0.075cm); 
\draw[thick, fill=white] (-1.25,-0.5) circle (0.075cm); 
\Text[x=-.75,y=0.0, anchor = center]{$=$}
\draw[thick] (-.25,0.5) -- (.75,.5);
\draw[thick] (-.25,-0.5)-- (.75,-0.5);
\draw[thick, fill=white] ( .75, 0.5) circle (0.075cm); 
\draw[thick, fill=white] ( .75,-0.5) circle (0.075cm); 
\end{tikzpicture} \, .
\label{eq:swap_diag}
\end{split}
\ee
This verifies the unitality of $\hat{K}_s$, but also that it is a dual-Koopman operator. It is easy to see that it is not $\Gamma$-dual or perfect, and that if $\hat{K}$ is dual ($\Gamma$-dual), then $\hat{K}_s \hat{K}$ and  $\hat{K} \hat{K}_s$ are $\Gamma$-dual (dual). Thus dual-Koopman operators can be constructed by composition of a $\Gamma$-dual operator with the swap. 

Let $V_{12}(q_1,q_2)$ be an arbitrary smooth interaction potential between the degrees of freedom. Consider the canonical transformation 
\beq
q_i'=q_i,\; p_i'=p_i -\partial V_{12}/\partial q_i,\; (i=1,2). 
\eeq
It is straightforward to verify that the associated Koopman operator, say $\hat{K}_{12}$,
satisfies Eq.~(\ref{eq:Gammadualunitality_diag}) and hence is $\Gamma$-dual. 
These are classical equivalents of the diagonal unitaries that are $\Gamma$-dual.
From this more general $\Gamma$-dual Koopman operators can be constructed using any local canonical transformations, for example as $(\hat{K}_1\otimes \hat{K}_2)\hat{K}_{12}$. 

A dynamical system we study in some detail below corresponds to  $V_{12}=-(b/4 \pi^2) \cos[2 \pi (q_1+q_2)]$ and the local transforms are standard maps. The standard map is a textbook example \cite{Ott_book} of two-dimensional area preserving maps or Poincar\'e surfaces of section that has a typical route to chaos for Hamiltonian systems through the breaking of KAM tori as integrability is broken. It is given by $\left(q_i'=q_i+p_i', p_i'=p_i-(\alpha_i/2 \pi) \sin(2 \pi q_i) \right)$,  where $i=1,2$ and $\alpha_i$ are the strength of the kicks. The individual standard maps are predominantly chaotic for $|\alpha_i|>5$, with the Lyapunov exponent that is $\approx \ln(|\alpha_i|/2)$, while they are near integrable for $|\alpha_i| \ll 1$.

The composed system with two standard maps and the interaction $V_{12}$ as above is a 4-dim. coupled standard map has been studied for long \cite{Froeschle_1972, Arul_2001, Richter_2014, Lange_2014}.
From the above discussion, the  corresponding Koopman operator is $\Gamma$-dual. 
Further composing it with the swap operation yields a dual partner whose Koopman operator $\hat{K}=\hat{K}_s (\hat{K}_1\otimes \hat{K}_2)\hat{K}_{12}$ is dual-Koopman. As a canonical transformation, this ``dual coupled standard map" is explicitly:
\begin{align}
&q_1'=q_2+p_1', q_2'=q_1+p_2'\\
&p_1'=p_2-(\alpha_2/2 \pi)\sin(2\pi q_2)-(b/2 \pi) \sin[2 \pi(q_1+q_2)] \nonumber\\
&p_2'=p_1-(\alpha_1/2 \pi)\sin(2\pi q_1)-(b/2 \pi) \sin[2 \pi(q_1+q_2)].\nonumber
\label{eq:4DdualStandardMap}
\end{align}

Restricting the standard maps to the unit two-torus $T$, the coupled dynamical system is on $T \times T$. Explicit forms of the Koopman operator can be obtained by using the Fourier basis to represent functions:
$\braket{q,p|F_{m,n}}\equiv \braket{q,p|m,n}=\exp[2 \pi i (qm+pn)]$, $m,n=0, \pm 1, \cdots$. The uniform function $F_u$ is simply $F_{0,0}$.
In this basis, with $i=1,2$, the Koopman operators are:
\begin{align}
&\bra{m_i',n_i'}\hat{K}_i\ket{m_i,n_i}=\text{J}_{m_i-m_i'}(\alpha_i n_i') \, \delta_{n_i',m_i+n_i},\\
& \bra{m_1',n_1';m_2',n_2'}\hat{K}_{12}\ket{m_1,n_1;m_2,n_2}= \nonumber\\\
&\delta_{n_1,n_1'}\delta_{n_2,n_2'}\delta_{m_1-m_1',m_2-m_2'}\, \text{J}_{m_1-m_1'}[b(n_1+n_2)],
\label{eq:Koopman_SM}
\end{align}
where $\text{J}_{\ell}(x)$ is the Bessel function of the first kind.
In terms of this basis, we note the condition for unitality is that $\hat{K}$ have 
$\ket{0000}$ as eigenvector with eigenvalue $1$, while for duality $\hat{K}^R$ and for $\Gamma$-duality $\hat{K}^{\Gamma}$ must have this property. Here (with $\ell$ denoting the pairs $(m,n)$) $\bra{\ell_1'\ell_2'}\hat{K}\ket{\ell_1\ell_2}=
\bra{\ell_1'\ell_1}\hat{K}^R\ket{\ell_2'\ell_2}=
\bra{\ell_1'\ell_2}\hat{K}^{\Gamma}\ket{\ell_1\ell_2'}=$ 
are realignments and partial transposes that play a crucial role in the quantum context as well \cite{ASA_2021}.

\begin{figure*}
\centering
\includegraphics[scale=.3]{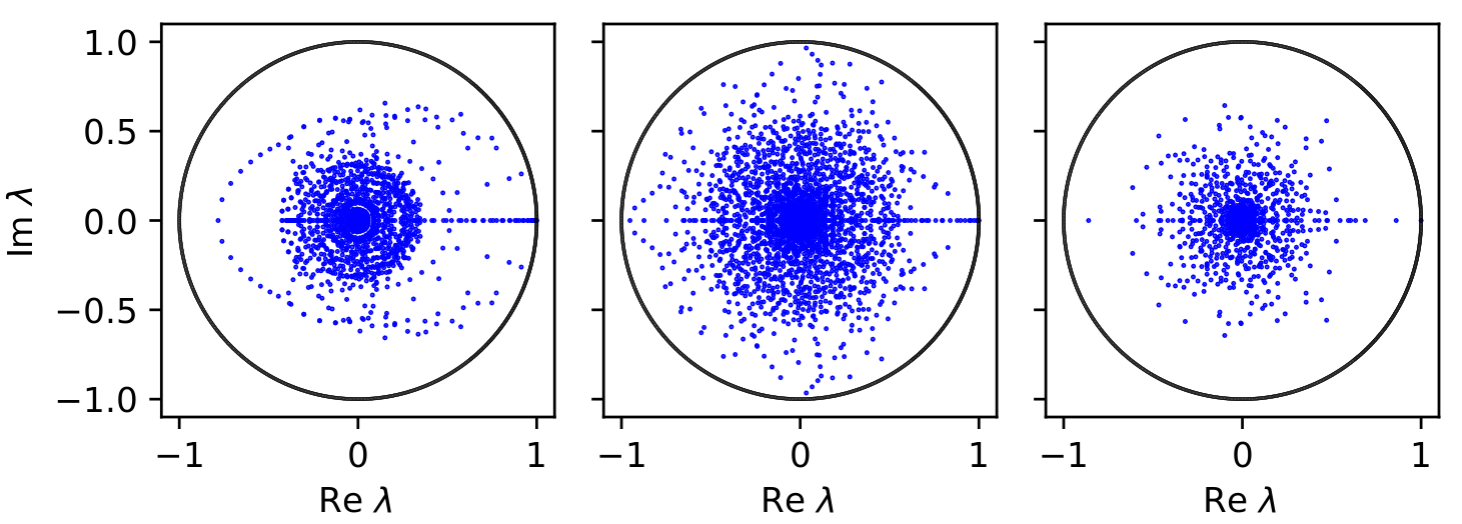}
\caption{The (real and imaginary parts of) eigenvalues of the Koopman map $\hat{K}^+_{\Gamma}$ determining the mixing rates of the dual-standard map circuit are shown for the cases (left to right) when $\alpha_1=0.1,\,2.0,\, 6.0$ and $b=0.1$.
A total of $91\times91$ modes are used for truncation.}
\label{fig:Resonances}
\end{figure*}

We now construct many-body systems as in the quantum brick-wall case,
with $2L$ particles at integer and half-integer sites $x=0,\, \nicefrac{1}{2},\, 1, \cdots, L-\nicefrac{1}{2}$, with periodic boundary conditions. Let $\hat{K}_{x,x+\nicefrac{1}{2}}$ be the Koopman operator that connects sites $x$ and $x+\nicefrac{1}{2}$. The corresponding many-body Koopman operator is
\beq
\hat{\mathbb K}= \bigotimes_{k=0}^{L-1} \hat{K}_{k-\nicefrac{1}{2},k} \bigotimes_{k=0}^{L-1} \hat{K}_{k,k+\nicefrac{1}{2}}.
\eeq
If $f_0(\xi_0)$ be a localized observable on particle $0$, it 
evolves after time $t=0,1,2,\cdots$ to $\hat{\mathbb K}^t f_0(\xi_0) F_u(\xi_{1/2})\cdots F_u(\xi_{L-1/2})$, which is generally scrambled across the particles. The basic correlation function of interest is
\beq
C(x,t)=\int_{\Omega} f_x(\xi_x) \hat{\mathbb K}^t f_0(\xi_0) \, d\xi,
\label{eq:Correlation}
\eeq
where we have dropped the uniform functions and the integration is over all the $2L$ particle phase-spaces. It is sufficient to consider the observables $f(\xi)$
to have zero mean, that is orthogonal to $F_u(\xi)$. We consider the case with translation symmetry when all the $K_{x,x+\nicefrac{1}{2}}$ are identical. This can be represented by a tensor network, and in Eq.~(\ref{eq:Correlation_diag}) this {\blue is}
shown for the special case of $2L=8$, eight particles, and $t=2$ and $x=\nicefrac{3}{2}$. $C(x,t)=$
\be
\begin{tikzpicture}[baseline=(current bounding box.center), scale=.7]
\def\eps{0};
\def\shift{11}
\def\shifty{-2.5}
\Text[x=\shift-3, y=-4]{}

\foreach \jj[evaluate=\jj as \j using -2*(ceil(\jj/2)-\jj/2)] in {0,...,3}
\foreach \i in {2,...,5}
{
    \draw[very thick] (\shift+.5-2*\i-1*\j,2+1*\jj+\shifty) -- (\shift+1-2*\i-1*\j,1.5+\jj+\shifty);
    \draw[very thick] (\shift+1-2*\i-1*\j,1.5+1*\jj+\shifty) -- (\shift+1.5-2*\i-1*\j,2+\jj+\shifty);
}

\foreach \jj[evaluate=\jj as \j using -2*(ceil(\jj/2)-\jj/2)] in {0,...,3}
\foreach \i in {2,...,5}
{
    \draw[very thick] (\shift+.5-2*\i-1*\j,1+1*\jj+\shifty) -- (\shift+1-2*\i-1*\j,1.5+\jj+\shifty);
    \draw[very thick] (\shift+1-2*\i-1*\j,1.5+1*\jj+\shifty) -- (\shift+1.5-2*\i-1*\j,1+\jj+\shifty);
    \draw[thick, fill=myblue, rounded corners=1pt] (\shift+0.75-2*\i-1*\j,1.75+\jj+\shifty) rectangle (\shift+1.25-2*\i-1*\j,1.25+\jj+\shifty);
    \draw[thick] (\shift+1-2*\i-1*\j,1.65+1*\jj+\shifty) -- (\shift+1.15-2*\i-1*\j,1.65+1*\jj+\shifty) -- (\shift+1.15-2*\i-1*\j,1.5+1*\jj+\shifty);
}
\draw[ thick, fill=black] (\shift-9.5,\shifty+1) circle (0.1cm); 
\draw[ thick, fill=white] (\shift-8.5,\shifty+1) circle (0.1cm);
\draw[ thick, fill=white] (\shift-7.5,\shifty+1) circle (0.1cm); 
\draw[ thick, fill=white] (\shift-6.5,\shifty+1) circle (0.1cm); 
\draw[ thick, fill=white] (\shift-5.5,\shifty+1) circle (0.1cm); 
\draw[ thick, fill=white] (\shift-4.5,\shifty+1) circle (0.1cm); 
\draw[ thick, fill=white] (\shift-3.5,\shifty+1) circle (0.1cm); 
\draw[ thick, fill=white] (\shift-2.5,\shifty+1) circle (0.1cm);
\draw[ thick, fill=white] (\shift-8.5,\shifty+5) circle (0.1cm); 
\draw[ thick, fill=white] (\shift-7.5,\shifty+5) circle (0.1cm);
\draw[ thick, fill=black] (\shift-6.5,\shifty+5) circle (0.1cm); 
\draw[ thick, fill=white] (\shift-5.5,\shifty+5) circle (0.1cm); 
\draw[ thick, fill=white] (\shift-4.5,\shifty+5) circle (0.1cm); 
\draw[ thick, fill=white] (\shift-3.5,\shifty+5) circle (0.1cm); 
\draw[ thick, fill=white] (\shift-2.5,\shifty+5) circle (0.1cm); 
\draw[ thick, fill=white] (\shift-1.5,\shifty+5) circle (0.1cm); 
\Text[x=-9.5+\shift,y=.5+\shifty]{$0$}
\Text[x=-8.5+\shift,y=.5+\shifty]{$\nicefrac{1}{2}$}
\Text[x=-7.5+\shift,y=.5+\shifty]{$1$}
\Text[x=-6.5+\shift,y=.5+\shifty]{$\nicefrac{3}{2}$}
\Text[x=-5.5+\shift,y=.5+\shifty]{$2$}
\Text[x=-4.5+\shift,y=.5+\shifty]{$\nicefrac{5}{2}$}
\Text[x=-3.5+\shift,y=.5+\shifty]{$3$}
\Text[x=-2.5+\shift,y=.5+\shifty]{$\nicefrac{7}{2}$}

\Text[x=-1.5+\shift,y=5.5+\shifty]{$0$}
\Text[x=-9+\shift,y=5.5+\shifty]{$\nicefrac{1}{2}$}
\Text[x=-7.5+\shift,y=5.5+\shifty]{$1$}
\Text[x=-6.5+\shift,y=5.5+\shifty]{$\nicefrac{3}{2}$}
\Text[x=-5.5+\shift,y=5.5+\shifty]{$2$}
\Text[x=-4.5+\shift,y=5.5+\shifty]{$\nicefrac{5}{2}$}
\Text[x=-3.5+\shift,y=5.5+\shifty]{$3$}
\Text[x=-2.5+\shift,y=5.5+\shifty]{$\nicefrac{7}{2}$}

\Text[x=-10.25+\shift,y=1.5+\shifty]{$\nicefrac{1}{2}$}
\Text[x=-10.25+\shift,y=2.5+\shifty]{$1$}
\Text[x=-10.25+\shift,y=3.5+\shifty]{$\nicefrac{3}{2}$}
\Text[x=-10.25+\shift,y=4.5+\shifty]{$2$}

\draw[->] (-6.5+\shift,-0.5+\shifty) -- (-5.5+ \shift,-0.5+\shifty);
\Text[x=-6.75+\shift,y=-0.5+\shifty]{$x$}
\draw[->] (-11.25+\shift,1.75+\shifty) -- (-11.25+ \shift,2.75+\shifty);
\node[rotate=90] at (-11.25+\shift,1.5+\shifty){$t$};
\end{tikzpicture}
\label{eq:Correlation_diag}
\ee
Using the unitality diagrams in Eq.~(\ref{eq:unitality_diag}) simplifies the diagram considerably already and we are left with a rectangle that includes only the sites $\nicefrac{3}{2},2$ at $t=2$ with $0, \nicefrac{1}{2}$ at $t=0$, the uniform functions moving in to the wires connecting the rectangle. In this example, the ``rectangle" is only one gate wide, but in general the width depends on the spatial separation between the two filled circles.

Using the dual-unitality diagrams in Eq.~(\ref{eq:dualunitality_diag}) now simplifies this further and renders it $0$. In fact, with the initial function localized at site $0$,  as the function at time $t=2$ varies over the sites, all the correlations vanish except the case $x=2$, when there is a direct light cone connecting the ``occupied" sites. As the rules are identical to the quantum case, more elaborate discussions are available in \cite{Kos_2021}.
The main result is that $C(x=t,t)$  is the only nonzero correlation for $t \leq L/2$, if the initial observable was in an integer site. If it was in a half-integer site the correlation propagates to the left, hence the only non-vanishing correlations for dual-Koopman circuits are on the light cone $C_{\pm}(\pm t,t)$ and they vanish everywhere inside it for these times. On the light cone, the correlations are determined simply by the $\Gamma$ maps of Eq.~(\ref{eq:Maps}):
\beq
C_+(t,t)=\bra{f_t}\left(\hat{K}_{\Gamma}^{+}\right)^{2t}\ket{f_0},
\label{eq:Correl_Maps1}
\eeq
and $\hat{K}_{\Gamma}^{-}$ governs the left moving correlation $C_{-}(-t,t)$.

Thus the eigenvalues of $\hat{K}^{\pm}_{\Gamma}$ derived from the dual-Koopman operators determine the decay of correlations in the many-body circuits. In the thermodynamic limit, they are all that determine the mixing. While the Ruelle-Pollicot resonances \cite{Ruelle1986} that determine correlation decay are difficult to find even in simple dynamical systems, such as a single standard map, it is quite remarkable how for dual-Koopman circuits they are accessible so simply.

Returning to the example of the dual coupled standard map in Eq.~(\ref{eq:4DdualStandardMap}), we use it to construct many-body coupled standard maps on $2L$ copies of the 2-torus. In the Fourier basis we get 
\beq
\begin{split}
&\bra{m',n'}\hat{K}^+_{\Gamma}\ket{m,n}=\bra{0,0;m',n'}\hat{K}\ket{m,n;0,0}\\
& = \text{J}_0(b\,n) \, \delta_{n',m+n}  \, \text{J}_{m-m'}(\alpha_1n'). 
\end{split}
\eeq
The matrix elements are evaluated using the Koopman operators in Eq.~(\ref{eq:Koopman_SM}) and take this extraordinarily simple form, 
with the Bessel function of order $0$ being the only distinguishing 
feature from this channel and the unitary Koopman operator for a single standard map. Note that $\hat{K}^+_{\Gamma}$ does not depend on $\alpha_2$, while $\hat{K}^{-}_{\Gamma}$ does not depend on $\alpha_1$. Let the eigenvalues of $\hat{K}^+_{\Gamma}$ be $\lambda_0, \lambda_1, \cdots$ arranged in a descending order of their magnitude, with $\lambda_0=1$ being the trivial eigenvalue.
They determine the mixing rates and if $|\lambda_1|<1$ all the modes are mixing, leading to thermalization. 

As an extreme case, when $\alpha_1=\alpha_2=0$ the uncoupled maps are integrable, for $b=0$, the only interaction is provided by the swap and the many-body system can also be considered integrable. For $b\neq 0$, $\bra{m',n'}\hat{K}^+_{\Gamma}\ket{m,n}=\delta_{m,m'}\delta_{n',m+n}\text{J}_0(bn)$. Numerical results indicate that even for small $b$, 
$|\lambda_1|<1$, implying that in the thermodynamic limit, mixing occurs with arbitrarily small interaction. It is easy to verify that $\braket{m,n|\Phi_{k}}=\delta_{m,0}\delta_{n,k}$ is an eigenmode with eigenvalue $\text{J}_0(bk)$ for $k=0,1,\cdots$, and the numerical values match, especially $|\lambda_1|=\text{J}_0(b)\approx 1-b^2/4 <1$, indicating mixing. In \cite{Bertini_2021CMP} it is shown that in the quantum case the spectral form factor matches with the random matrix value if the gates of the dual circuit differed from the swap. This is consistent with the classical limit being fully mixing even arbitrarily close to swap.

For a generic case, Fig.~\ref{fig:Resonances} shows how the resonances or eigenvalues $\lambda_i$ of the Koopman channel $\hat{K}_{\Gamma}^+$, behave as the single particle chaos parameter is changed and the largest nontrivial eigenvalue moves in. Most of the eigenvalues not clustering at the origin seem converged as verified by changing the truncation size. Interestingly the largest nontrivial eigenavalue determining the mixing rate is real and is seen to be well approximated for even moderately large  $\alpha_1$ to be $J_0(b)$.  More detailed studies are left for the future.

The above class of dual-Koopman operators cannot be ``perfect". However we can find  examples of these as well. A coupled cat map lattice was mentioned in \cite{ASA_2021}
which turns out to be a simple example of such a system. Let
\beq
\begin{bmatrix}
q_1'\\p'_1\\q_2'\\p_2'
\end{bmatrix}
=
\begin{bmatrix}
1 & 0&0&1\\0&2&1&0\\
0&1&1&0\\
1&0&0&2 
\end{bmatrix}
\begin{bmatrix}
q_1\\p_1\\q_2\\p_2
\end{bmatrix} \text{mod\; 1},
\label{eq:4DcatmapClassical}
\eeq
be a canonical transformation. This is a pair of Arnold cat maps \cite{Arnold_Avez} on the $(q_1,p_2)$ 
and $(q_2,p_1)$ planes. The Koopman operator is  $\bra{m_1',n_1';m_2',n_2'}\hat{K}_{Cat}\ket{m_1,n_1;m_2,n_2}=\delta_{m_1',m_1+n_2}\delta_{n_1',2n_1+m_2}\delta_{m_2',n_1+m_2}\delta_{n_2',m_1+2n_2}$. It is straightforward to check that it is unital and so are the realigned and partial transposed versions. Hence this Koopman is perfect and the resulting single particle correlations in a brick wall circuit of these are Bernoulli, that is $C(x=0,t=0)$ is the only non-vanishing correlation, else everywhere, including on the light cone, the correlations vanish. Note that any local canonical  transformations, including nonlinear maps, can be used to define a host of interesting dynamical systems whose corresponding circuits are all Bernoulli.

Of further interest is quantum-classical correspondence in dual-unitary and dual-Koopman circuits. Both the coupled standard maps discussed above and the cat maps have natural quantizations and we expect that beyond the time $t=L/2$ while the quantum correlations recur and oscillate, the classical correlations can continue to decay. 

[{\sl Note: On completion of the work, a closely related work has appeared \cite{Prosen_2023} based on the adjoint of the Koopman operators used in this work. Conditions on the underlying canonical transform to result in dual unitality of the adjoint Perron-Frobenius operator are presented there.}]

\begin{acknowledgments}
I am grateful to the organizers of the program ``Dynamical Foundations
of Many-Body Quantum Chaos" at the Institut Pascal, Universit\'e Paris-Saclay, March-April 2023, where the format enabled this work to be initiated.

\end{acknowledgments}


\end{document}